\documentclass[journal,twoside,web]{ieeecolor}
\usepackage{generic}
\usepackage{cite}
\usepackage{amsmath,amssymb,amsfonts}
\usepackage{algorithmic}
\usepackage{graphicx}
\usepackage{algorithm,algorithmic}
\usepackage{hyperref}
\hypersetup{hidelinks=true}
\usepackage{textcomp}
\def\BibTeX{{\rm B\kern-.05em{\sc i\kern-.025em b}\kern-.08em
    T\kern-.1667em\lower.7ex\hbox{E}\kern-.125emX}}
\begin{document}
\title{Quaternion MLP Neural Networks Based on the Maximum Correntropy Criterion}
\author{Gang Wang*, Xinyu Tian, and Zuxuan Zhang
\thanks{Gang Wang, Xinyu Tian, and Zuxuan Zhang are with the Center for Robotics, School of Information and Communication Engineering, University of Electronic Science and Technology of China, Chengdu 611731, P.R. China (e-mail: wanggang\_hld@uestc.edu.cn, 1594048742@qq.com, changtsuhsuan@qq.com).}
}

\maketitle

\begin{abstract}
	We propose a gradient ascent algorithm for quaternion multilayer perceptron (MLP) networks based on the cost function of the maximum correntropy criterion (MCC). In the algorithm, we use the split quaternion activation function based on the generalized Hamilton-real quaternion gradient. By introducing a new quaternion operator, we first rewrite the early quaternion single layer perceptron algorithm. Secondly, we propose a gradient descent algorithm for quaternion multilayer perceptron based on the cost function of the mean square error (MSE). Finally, the MSE algorithm is extended to the MCC algorithm. Simulations show the feasibility of the proposed method.
\end{abstract}


\section{Introduction}
\label{sec:introduction}
Tultilayer (MLP) neural networks are the most popular neural networks. The connection weights can be adjusted to represent the input-output relations. Recently MLP neural network models have been explored based on quaternions in \cite{bib1,bib2,bib3,bib4,bib5,bib6}, because the 4-D signals in the quaternion domain can incorporate mutual information to reduce complexity. Applications of quaternion neural networks include image classification \cite{bib7,bib8,bib9}, signal classification \cite{bib10,bib11}, image retrieval \cite{bib12}, the control problem \cite{bib13,bib14}, the prediction of time series \cite{bib15,bib16,bib17}, forecasting three-dimensional wind signals \cite{bib35}, pose estimation \cite{bib18,bib19} and convolutional neural networks \cite{bib20,bib21,bib22}.
The adjusting process of quaternion connection weights is also called learning. Quaternion learning algorithms can be loosely divided into not-gradient-based methods, such as extreme learning machines \cite{bib23,bib24,bib25}, and gradient-based methods, such as least-mean-square (LMS) algorithms \cite{bib26}. This paper is focused on gradient-based methods.
When the LMS algorithm, or the back-propagation algorithm, of MLP neural networks was generalized from the real domain 
to the complex domain, the analyticity of the activation functions had to be studied \cite{bib27,bib28,bib29}. If an activation function is analytic, then it is a constant function. As an alternative, a split complex activation function was proposed for the MLP. Following the split idea in the complex domain, a split quaternion activation function was proposed in and used in a nonlinear filter, or single layer perceptron(SLP). In addition to the split function, in and, another activation function was introduced using a local analyticity condition for a quaternion nonlinear filter and echo state networks, respectively. 
However, the quaternion gradient in misused the non-commutative property, and it prevented the extension of the quaternion SLP to the MLP. From 2014 to 2019, the correct quaternion gradient was proposed using three approaches: generalized Hamilton-real (GHR) calculus \cite{bib30,bib31,bib32}, the quaternion product \cite{bib33}, and the quaternion involutions \cite{bib34}. Hitzer detected the gradient mistake in, and obtained corrections based on Clifford calculus.  
The LMS algorithm is based on the cost function of the mean square error (MSE), and works well in Gaussian additive noise, whose higher statistics are constants. In real scenarios, non-Gaussian problems arise \cite{bib36}. The signals are often contaminated by non-Gaussian noise or impulsive noise. The performance of the above LMS algorithm may be poor in non-Gaussian noise.
Recently, information theoretic learning was proposed for non-Gaussian noise. The proposed cost functions include \cite{bib12} the maximum correntropy criterion (MCC), improved least sum of exponentials, and least mean kurtosis. The MCC is robust to large outliers or impulsive noise, and has been applied in quaternion filters. This motivates us to study quaternion MLP networks based on the MCC \cite{bib18,bib19}.
In this paper, we revisit the split quaternion activation function based on the GHR quaternion gradient, and rewrite the quaternion nonlinear filter algorithm by introducing a new quaternion operator. Then we propose two LMS algorithms for the quaternion MLP based on the MSE and MCC. Simulations show the feasibility of the proposed method\cite{bib37,bib38}.

\section{Preliminaries}
\subsection{Quaternion Algebra }
The quaternion domain is a non-commutative extension of the complex domain. A quaternion variable   consists of a real part and three imaginary components and can be expressed as

\[{\mathop{\rm q}\nolimits}  = {{\mathop{\rm q}\nolimits} _{\mathop{\rm a}\nolimits} } + \iota {{\mathop{\rm q}\nolimits} _{\mathop{\rm b}\nolimits} } + J{{\mathop{\rm q}\nolimits} _{\mathop{\rm c}\nolimits} } + \kappa {{\mathop{\rm q}\nolimits} _{\mathop{\rm d}\nolimits} }\]

The quaternion product is non-commutative, that is,  but   Given   we have
\[\left\{ {\begin{array}{*{20}{c}}
		{\iota J = \kappa {\rm{,  }}J\kappa  = \iota {\rm{,  }}\kappa \iota  = J{\rm{,}}}\\
		{\iota J\kappa  = {\iota ^2} = {J^2} = {\kappa ^2} =  - {\rm{1}}{\rm{.}}}
\end{array}} \right.\]
\[{{\mathop{\rm q}\nolimits} _1}{{\mathop{\rm q}\nolimits} _2} \ne {{\mathop{\rm q}\nolimits} _2}{{\mathop{\rm q}\nolimits} _1}.\]
Three perpendicular quaternion involutions are given by
\[\left\{ {\begin{array}{*{20}{c}}
		{{{\mathop{\rm q}\nolimits} ^\iota } =  - \iota {\mathop{\rm q}\nolimits} \iota  = {{\mathop{\rm q}\nolimits} _{\mathop{\rm a}\nolimits} } + \iota {{\mathop{\rm q}\nolimits} _{\mathop{\rm b}\nolimits} } - J{{\mathop{\rm q}\nolimits} _{\mathop{\rm c}\nolimits} } - \kappa {{\mathop{\rm q}\nolimits} _{\mathop{\rm d}\nolimits} }{\rm{,}}}\\
		{{{\mathop{\rm q}\nolimits} ^J} =  - J{\mathop{\rm q}\nolimits} J = {{\mathop{\rm q}\nolimits} _{\mathop{\rm a}\nolimits} } - \iota {{\mathop{\rm q}\nolimits} _{\mathop{\rm b}\nolimits} } + J{{\mathop{\rm q}\nolimits} _{\mathop{\rm c}\nolimits} } - \kappa {{\mathop{\rm q}\nolimits} _{\mathop{\rm d}\nolimits} }{\rm{,}}}\\
		{{{\mathop{\rm q}\nolimits} ^\kappa } =  - \kappa {\mathop{\rm q}\nolimits} \kappa  = {{\mathop{\rm q}\nolimits} _{\mathop{\rm a}\nolimits} } - \iota {{\mathop{\rm q}\nolimits} _{\mathop{\rm b}\nolimits} } - J{{\mathop{\rm q}\nolimits} _{\mathop{\rm c}\nolimits} } + \kappa {{\mathop{\rm q}\nolimits} _{\mathop{\rm d}\nolimits} }.}
\end{array}} \right.\]
Then the quaternion conjugate operation can be expressed as
\[{{\mathop{\rm q}\nolimits} ^ * } = {{\mathop{\rm q}\nolimits} _{\mathop{\rm a}\nolimits} } - \iota {{\mathop{\rm q}\nolimits} _{\mathop{\rm b}\nolimits} } - J{{\mathop{\rm q}\nolimits} _{\mathop{\rm c}\nolimits} } - \kappa {{\mathop{\rm q}\nolimits} _{\mathop{\rm d}\nolimits} } = \frac{1}{2}\left( {{{\mathop{\rm q}\nolimits} ^\iota } + {{\mathop{\rm q}\nolimits} ^J} + {{\mathop{\rm q}\nolimits} ^\kappa } - {\mathop{\rm q}\nolimits} } \right).\]
The norm of a quaternion is defined as
\[\left\| {\mathop{\rm q}\nolimits}  \right\|_{^{\rm{2}}}^{\rm{2}} = {\mathop{\rm q}\nolimits} {{\mathop{\rm q}\nolimits} ^ * } = {\mathop{\rm q}\nolimits} _{\mathop{\rm a}\nolimits} ^{\rm{2}} + {\mathop{\rm q}\nolimits} _{\mathop{\rm b}\nolimits} ^{\rm{2}} + {\mathop{\rm q}\nolimits} _{\mathop{\rm c}\nolimits} ^{\rm{2}} + {\mathop{\rm q}\nolimits} _{\mathop{\rm d}\nolimits} ^{\rm{2}}.\]

$HR$ calculus is an extension of Wirtinger calculus, which is also called   calculus.   calculus provides a simple and straightforward approach to calculating derivatives with respect to complex parameters. HR calculus comprises the HR derivatives
\[{\kern 1pt} {\kern 1pt} \frac{{\partial J}}{{\partial {{\bf{q}}^ * }}} = \frac{1}{4}\left( {\frac{{\partial J}}{{\partial {{\bf{q}}_a}}} + \frac{{\partial J}}{{\partial {{\bf{q}}_b}}}\iota  + \frac{{\partial J}}{{\partial {{\bf{q}}_c}}}J + \frac{{\partial J}}{{\partial {{\bf{q}}_d}}}\kappa } \right).\]

\subsection{Quaternion Split Product}
Definition 1. (Quaternion Split Product ) For two quaternions, $x, y$ , the quaternion split product is defined as
\[x \odot y = y \odot x = {x_{\mathop{\rm a}\nolimits} }{y_{\mathop{\rm a}\nolimits} } + \iota {x_{\mathop{\rm b}\nolimits} }{y_{\mathop{\rm b}\nolimits} } + J{x_{\mathop{\rm c}\nolimits} }{y_{\mathop{\rm c}\nolimits} } + \kappa {x_{\mathop{\rm d}\nolimits} }{y_{\mathop{\rm d}\nolimits} } \in {H^{\mathop{\rm m}\nolimits} }{\rm{,}}\]
where   represents the dot product. For two quaternions,  , the quaternion split product is defined as
\[{\mathop{\rm x}\nolimits}  \odot y = y \odot {\mathop{\rm x}\nolimits}  = {{\mathop{\rm x}\nolimits} _{\mathop{\rm a}\nolimits} }{y_{\mathop{\rm a}\nolimits} } + \iota {{\mathop{\rm x}\nolimits} _{\mathop{\rm b}\nolimits} }{y_{\mathop{\rm b}\nolimits} } + J{{\mathop{\rm x}\nolimits} _{\mathop{\rm c}\nolimits} }{y_{\mathop{\rm c}\nolimits} } + \kappa {{\mathop{\rm x}\nolimits} _{\mathop{\rm d}\nolimits} }{y_{\mathop{\rm d}\nolimits} } \in {H^{\mathop{\rm m}\nolimits} }\]

The quaternion split product has the following property,
\[{\left( {x \odot y} \right)^{\rm{*}}} = x \odot {y^{\rm{*}}} = {x^{\rm{*}}} \odot y = {y^{\rm{*}}} \odot x = y \odot {x^{\rm{*}}}{\rm{.}}\]

\subsection{Nonlinear Filtering or SLP}
When the quaternion nonlinear filtering problem is considered in the discrete time domain i, there is an input quaternion vector ${\bf{u}}\left( i \right) \in {H^n}$  with the unknown original quaternion parameter  ${{\bf{w}}^o} \in {H^n}$ and the desired response $d\left( i \right) \in {H^1}$  as follows:
\[\left\{ \begin{array}{l}
	{\mathop{\rm x}\nolimits}  = {{\bf{w}}^H}u\left( {\mathop{\rm i}\nolimits}  \right){\rm{,}}\\
	{\mathop{\rm x}\nolimits}  = {{\mathop{\rm x}\nolimits} _{\mathop{\rm a}\nolimits} } + \iota {{\mathop{\rm x}\nolimits} _{\mathop{\rm b}\nolimits} } + J{{\mathop{\rm x}\nolimits} _{\mathop{\rm c}\nolimits} } + \kappa {{\mathop{\rm x}\nolimits} _{\mathop{\rm d}\nolimits} } \in {H^1}{\rm{,}}\\
	{\bf{w}} = {{\bf{w}}_{\mathop{\rm a}\nolimits} } + \iota {{\bf{w}}_{\mathop{\rm b}\nolimits} } + J{{\bf{w}}_{\mathop{\rm c}\nolimits} } + \kappa {{\bf{w}}_{\mathop{\rm d}\nolimits} } \in {H^n}{\rm{,}}\\
	u\left( {\mathop{\rm i}\nolimits}  \right) = {u_{\mathop{\rm a}\nolimits} } + \iota {u_{\mathop{\rm b}\nolimits} } + J{u_{\mathop{\rm c}\nolimits} } + \kappa {u_{\mathop{\rm d}\nolimits} } \in {H^n}.
\end{array} \right.\]

The error signal for the quaternion linear filter is defined as
\[e\left( i \right) = d\left( i \right) - \Phi \left[ {{\bf{w}}_{}^H{\bf{u}}\left( i \right)} \right].\]

The quaternion nonlinear function or the activation function is defined as
\[\Phi \left( x \right) = \phi \left( {{x_a}} \right) + \phi \left( {{x_b}} \right)\iota  + \phi \left( {{x_c}} \right)J + \phi \left( {{x_d}} \right)\kappa ,\]

where 
\[\left\{ \begin{array}{l}
	\phi \left( a \right) = {\rm{tanh}}\left( a \right),\\
	\frac{{\partial \phi \left( a \right)}}{{\partial a}} = {\rm{sec}}{{\rm{h}}^2}\left( a \right).
\end{array} \right.\]

The gradient of (13) is 
\[\begin{array}{c}
	{\rm{4}}\frac{{\partial \Phi }}{{\partial {x^ * }}} = {\rm{sec}}{{\rm{h}}^2}\left( {{x_a}} \right) + {\rm{sec}}{{\rm{h}}^2}\left( {{x_b}} \right)\iota \\
	+ {\rm{sec}}{{\rm{h}}^2}\left( {{x_c}} \right)J + {\rm{sec}}{{\rm{h}}^2}\left( {{x_d}} \right)\kappa .
\end{array}\]

The cost function of the MSE is defined as  to estimate the parameter  
\[J({\bf{w}}) = e\left( i \right)\left( {d_{}^ * \left( i \right) - \Phi _{}^ * \left[ x \right]} \right){\rm{ = }}\left( {d\left( i \right) - \Phi \left[ x \right]} \right)e_{}^ * \left( i \right).\]

The corresponding gradient is defined as
\[{\kern 1pt} {\kern 1pt} \frac{{\partial J}}{{\partial {{\bf{w}}^ * }}} = \frac{1}{4}\left( {\frac{{\partial J}}{{\partial {{\bf{w}}_a}}} + \frac{{\partial J}}{{\partial {{\bf{w}}_b}}}\iota  + \frac{{\partial J}}{{\partial {{\bf{w}}_c}}}J + \frac{{\partial J}}{{\partial {{\bf{w}}_d}}}\kappa } \right).\]

The quaternion gradient descent algorithm was proposed in [35] 
\[\begin{array}{l}
	{\kern 1pt} {\kern 1pt} {\bf{w}}\left( {i{\rm{ + 1}}} \right) = {\bf{w}}\left( i \right) + \eta {\bf{u}}\left( i \right){e^*}\left( i \right) = {\bf{w}}\left( i \right) + \\
	\eta {\bf{u}}\left[ {{\rm{sec}}{{\rm{h}}^2}\left( {{x_a}} \right){e_a} - {\rm{sec}}{{\rm{h}}^2}\left( {{x_b}} \right){e_b}\iota  - {\rm{sec}}{{\rm{h}}^2}\left( {{x_c}} \right){e_c}J - {\rm{sec}}{{\rm{h}}^2}\left( {{x_d}} \right){e_d}\kappa } \right],
\end{array}\]

where is the step size. Using the definition of the quaternion split product (8), we rewrite it as
\[{\kern 1pt} {\kern 1pt} {\bf{w}}\left( {i{\rm{ + 1}}} \right) = {\bf{w}}\left( i \right) + \eta {\bf{u}}\left( i \right)\left[ {{\kern 1pt} {\kern 1pt} \frac{{\partial \Phi }}{{\partial {x^ * }}} \odot {e^ * }\left( i \right)} \right],\]

which has a similar expression to that in the complex domain.

\section{Quaternion MLP Based on The MSE}
Fig. 1 shows the schematic diagram of the MLP. The following expression is used: 

\begin{figure}[!t]
	\includegraphics[width=\columnwidth]{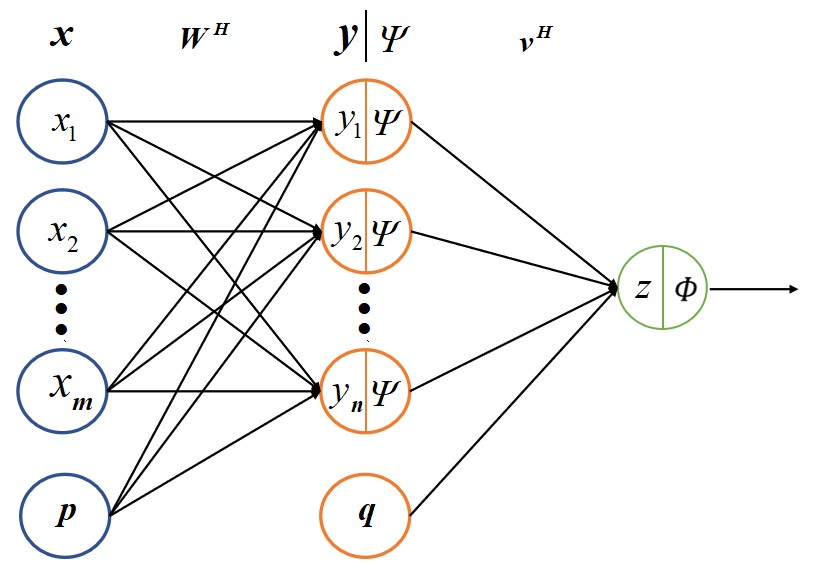}
	\caption{ Schematic diagram of  the MLP}
	\label{fig1}
\end{figure}

\[\left\{ \begin{array}{l}
	y = {W^{\mathop{\rm H}\nolimits} }x + p{\rm{,}}\\
	\Psi \left( y \right) = \Psi \left( {{W^{\mathop{\rm H}\nolimits} }x + p} \right){\rm{,}}\\
	{\mathop{\rm z}\nolimits}  = {v^H}\Psi \left( y \right) + {\mathop{\rm q}\nolimits} {\rm{,}}\\
	{h_{W,v}}\left( x \right) = \Phi \left( {\mathop{\rm z}\nolimits}  \right) = \Phi \left[ {{v^H}\Psi \left( y \right) + {\mathop{\rm q}\nolimits} } \right]{\rm{,}}\\
	= \Phi \left[ {{v^H}\Psi \left( {{W^{\mathop{\rm H}\nolimits} }x + p} \right) + {\mathop{\rm q}\nolimits} } \right]{\rm{,}}
\end{array} \right.\]
where

\[\left\{ \begin{array}{l}
	y = {\left[ {\begin{array}{*{20}{c}}
				{{\mathop{\rm y}\nolimits} _1^{}}&{{\mathop{\rm y}\nolimits} _2^{}}& \cdots &{{\mathop{\rm y}\nolimits} _m^{}}
		\end{array}} \right]^T}{\rm{,}}\Psi \left( y \right) \in {H^{\mathop{\rm n}\nolimits} }{\rm{,}}\\
	W = \left[ {\begin{array}{*{20}{c}}
			{w_1^{}}&{w_2^{}}& \cdots &{w_{\mathop{\rm n}\nolimits} ^{}}
	\end{array}} \right] \in {H^{{\mathop{\rm m}\nolimits}  \times {\mathop{\rm n}\nolimits} }}{\rm{,}}\\
	p = {\left[ {\begin{array}{*{20}{c}}
				{{\mathop{\rm p}\nolimits} _1^{}}&{{\mathop{\rm p}\nolimits} _2^{}}& \cdots &{{\mathop{\rm p}\nolimits} _m^{}}
		\end{array}} \right]^T} \in {H^{\mathop{\rm n}\nolimits} }{\rm{,}}\\
	v = {\left[ {\begin{array}{*{20}{c}}
				{{\mathop{\rm v}\nolimits} _1^{}}&{{\mathop{\rm v}\nolimits} _2^{}}& \cdots &{{\mathop{\rm v}\nolimits} _m^{}}
		\end{array}} \right]^T} \in {H^{\mathop{\rm n}\nolimits} }{\rm{,}}\\
	{\mathop{\rm y}\nolimits} _{\mathop{\rm i}\nolimits} ^{} = w_{\mathop{\rm i}\nolimits} ^{\mathop{\rm H}\nolimits} x + {\mathop{\rm p}\nolimits} _{\mathop{\rm i}\nolimits} ^{}{\rm{,}}{\mathop{\rm i}\nolimits}  = 1,2, \cdots {\mathop{\rm n}\nolimits} {\rm{,}}\\
	{\mathop{\rm z}\nolimits} {\rm{,}}q{\rm{,}}\Phi \left( z \right) \in {H^1}{\rm{, }}x \in {H^{\mathop{\rm m}\nolimits} }.
\end{array} \right.\]

The activation functions have the same expressions:
\[\left\{ \begin{array}{l}
	\Psi \left( {\bf{y}} \right) = \phi \left( {{{\bf{y}}_a}} \right) + \phi \left( {{{\bf{y}}_b}} \right)\iota  + \phi \left( {{{\bf{y}}_c}} \right)J + \phi \left( {{{\bf{y}}_d}} \right)\kappa ,\\
	\Phi \left( z \right) = \phi \left( {{z_a}} \right) + \phi \left( {{z_b}} \right)\iota  + \phi \left( {{z_c}} \right)J + \phi \left( {{z_d}} \right)\kappa .
\end{array} \right.\]

The cost function is defined as  to estimate the parameters W, V, p and q:

\[\left\{ \begin{array}{l}
	J\left( {{\bf{W}},{\bf{V}},{\bf{p}},q} \right) = ee_{}^ *  = e\left( {d_{}^ *  - \Phi _{}^ * \left[ z \right]} \right){\rm{ = }}\left( {d - \Phi \left[ z \right]} \right)e_{}^ * ,\\
	e = d - \Phi \left[ z \right].
\end{array} \right.\]

\subsection{Quaternion Gradient for V and q}
The gradients of v and q are similar to the gradient of w in the above nonlinear filter, and can be obtained by (18) directly as follows:

\[\begin{array}{c}
	- {\kern 1pt} {\kern 1pt} {\rm{4}}\frac{{\partial J}}{{\partial {q^ * }}} = {\rm{2}}\left[ {{\rm{sec}}{{\rm{h}}^2}\left( {{z_a}} \right){e_a} + {\rm{sec}}{{\rm{h}}^2}\left( {{z_b}} \right){e_b}\iota } \right.\\
	\left. { + {\rm{sec}}{{\rm{h}}^2}\left( {{z_c}} \right){e_c}J + {\rm{sec}}{{\rm{h}}^2}\left( {{z_d}} \right){e_d}\kappa } \right]\\
	= {\rm{2}}\frac{{\partial \Phi \left( z \right)}}{{\partial {z^ * }}} \odot e,
\end{array}\]

\[\begin{array}{c}
	- {\kern 1pt} {\kern 1pt} {\rm{4}}\frac{{\partial J}}{{\partial {{\bf{v}}^ * }}} = {\rm{2}}\Psi \left( {\bf{y}} \right)\left[ {{\rm{sec}}{{\rm{h}}^2}\left( {{z_a}} \right){e_a} - {\rm{sec}}{{\rm{h}}^2}\left( {{z_b}} \right){e_b}\iota } \right.\\
	\left. { - {\rm{sec}}{{\rm{h}}^2}\left( {{z_c}} \right){e_c}J - {\rm{sec}}{{\rm{h}}^2}\left( {{z_d}} \right){e_d}\kappa } \right]\\
	= {\rm{2}}\Psi \left( {\bf{y}} \right)\left[ {{\kern 1pt} {\kern 1pt} \frac{{\partial \Phi \left( z \right)}}{{\partial {z^ * }}} \odot {e^ * }} \right].
\end{array}\]

Those two equations above have similar expressions to those in the complex domain.

\subsection{Quaternion Gradient for p}

We rewrite it as follows to use the chain rule:

\[\left\{ \begin{array}{c}
	\left[ {\begin{array}{*{20}{c}}
			{{{\bf{y}}_a}}\\
			{{{\bf{y}}_b}}\\
			{{{\bf{y}}_c}}\\
			{{{\bf{y}}_d}}
	\end{array}} \right]{\rm{ = }}\left[ {\begin{array}{*{20}{c}}
			{{\bf{W}}_a^T}&{{\bf{W}}_b^T}&{{\bf{W}}_c^T}&{{\bf{W}}_d^T}\\
			{ - {\bf{W}}_b^T}&{{\bf{W}}_a^T}&{{\bf{W}}_d^T}&{ - {\bf{W}}_c^T}\\
			{ - {\bf{W}}_c^T}&{ - {\bf{W}}_d^T}&{{\bf{W}}_a^T}&{{\bf{W}}_b^T}\\
			{ - {\bf{W}}_d^T}&{{\bf{W}}_c^T}&{ - {\bf{W}}_b^T}&{{\bf{W}}_a^T}
	\end{array}} \right]\left[ {\begin{array}{*{20}{c}}
			{{\bf{x}}_a^{}}\\
			{{\bf{x}}_b^{}}\\
			{{\bf{x}}_c^{}}\\
			{{\bf{x}}_d^{}}
	\end{array}} \right] 
	\\+ \left[ {\begin{array}{*{20}{c}}
			{{{\bf{p}}_a}}\\
			{{{\bf{p}}_b}}\\
			{{{\bf{p}}_c}}\\
			{{{\bf{p}}_d}}
	\end{array}} \right],\\
	\Psi \left[ {\bf{y}} \right] = {\Psi _a} + {\Psi _b}\iota  + {\Psi _c}J + {\Psi _d}\kappa \\
	= \phi \left( {{{\bf{y}}_a}} \right) + \phi \left( {{{\bf{y}}_b}} \right)\iota  + \phi \left( {{{\bf{y}}_c}} \right)J + \phi \left( {{{\bf{y}}_d}} \right)\kappa ,\\
	\left[ {\begin{array}{*{20}{c}}
			{{z_a}}\\
			{{z_b}}\\
			{{z_c}}\\
			{{z_d}}
	\end{array}} \right]{\rm{ = }}\left[ {\begin{array}{*{20}{c}}
			{\Psi _a^T}&{\Psi _b^T}&{\Psi _c^T}&{\Psi _d^T}\\
			{\Psi _b^T}&{ - \Psi _a^T}&{ - \Psi _d^T}&{\Psi _c^T}\\
			{\Psi _c^T}&{\Psi _d^T}&{ - \Psi _a^T}&{ - \Psi _b^T}\\
			{\Psi _d^T}&{ - \Psi _c^T}&{\Psi _b^T}&{ - \Psi _a^T}
	\end{array}} \right]\left[ {\begin{array}{*{20}{c}}
			{{\bf{v}}_a^{}}\\
			{{\bf{v}}_b^{}}\\
			{{\bf{v}}_c^{}}\\
			{{\bf{v}}_d^{}}
	\end{array}} \right] \\
	+ \left[ {\begin{array}{*{20}{c}}
			{{q_a}}\\
			{{q_b}}\\
			{{q_c}}\\
			{{q_d}}
	\end{array}} \right],\\
	\Phi \left[ z \right] = \phi \left( {{z_a}} \right) + \phi \left( {{z_b}} \right)\iota  + \phi \left( {{z_c}} \right)J + \phi \left( {{z_d}} \right)\kappa .
\end{array} \right.\]

The corresponding gradient of p is defined as 
\[{\kern 1pt} {\kern 1pt} \frac{{\partial J}}{{\partial {{\bf{p}}^ * }}} = \frac{1}{4}\left( {\frac{{\partial J}}{{\partial {{\bf{p}}_a}}} + \frac{{\partial J}}{{\partial {{\bf{p}}_b}}}\iota  + \frac{{\partial J}}{{\partial {{\bf{p}}_c}}}J + \frac{{\partial J}}{{\partial {{\bf{p}}_d}}}\kappa } \right),\]

where
\[\left\{ \begin{array}{l}
	\frac{{\partial J}}{{\partial {{\bf{p}}_a}}} =  - \frac{{\partial e{\Phi ^ * }\left[ z \right]}}{{\partial {{\bf{p}}_a}}} - \frac{{\partial \Phi \left[ z \right]{e^ * }}}{{\partial {{\bf{p}}_a}}},\\
	\frac{{\partial J}}{{\partial {{\bf{p}}_b}}}\iota  =  - \frac{{\partial e{\Phi ^ * }\left[ z \right]}}{{\partial {{\bf{p}}_b}}}\iota  - \frac{{\partial \Phi \left[ z \right]{e^ * }}}{{\partial {{\bf{p}}_b}}}\iota ,\\
	\frac{{\partial J}}{{\partial {{\bf{p}}_c}}}J =  - \frac{{\partial e{\Phi ^ * }\left[ z \right]}}{{\partial {{\bf{p}}_c}}}J - \frac{{\partial \Phi \left[ z \right]{e^ * }}}{{\partial {{\bf{p}}_c}}}J,\\
	\frac{{\partial J}}{{\partial {{\bf{p}}_d}}}\kappa  =  - \frac{{\partial e{\Phi ^ * }\left[ z \right]}}{{\partial {{\bf{p}}_d}}}\kappa  - \frac{{\partial \Phi \left[ z \right]{e^ * }}}{{\partial {{\bf{p}}_d}}}\kappa .
\end{array} \right.\]

Note that
\[\frac{{\partial e{\Phi ^ * }\left[ z \right]}}{{\partial {{\bf{p}}_\delta }}} = {\left( {\frac{{\partial \Phi \left[ z \right]{e^ * }}}{{\partial {{\bf{p}}_\delta }}}} \right)^ * },\delta  \in \left\{ {a,b,c,d} \right\}.\left( {\begin{array}{*{20}{c}}
		1&0\\
		0&1
\end{array}} \right)\]

We only provide the details of ${{\partial e{\Phi ^ * }\left[ z \right]} \mathord{\left/
		{\vphantom {{\partial e{\Phi ^ * }\left[ z \right]} {\partial {{\bf{p}}_a}}}} \right.
		\kern-\nulldelimiterspace} {\partial {{\bf{p}}_a}}}$ , because of the limitation of space: 

\[\begin{array}{c}
	\frac{{\partial e{\Phi ^ * }\left[ z \right]}}{{\partial {{\bf{p}}_a}}} = e\left[ {\frac{{\partial \phi \left( {{z_a}} \right)}}{{\partial {{\bf{p}}_a}}} - \frac{{\partial \phi \left( {{z_b}} \right)}}{{\partial {{\bf{p}}_a}}}\iota  - \frac{{\partial \phi \left( {{z_c}} \right)}}{{\partial {{\bf{p}}_a}}}J - \frac{{\partial \phi \left( {{z_d}} \right)}}{{\partial {{\bf{w}}_a}}}\kappa } \right]\\
	= e\left[ {\frac{{\partial \phi \left( {{z_a}} \right)}}{{\partial {z_a}}}\frac{{\partial {z_a}}}{{\partial {\Psi _a}}}\frac{{\partial {\Psi _a}}}{{\partial {{\bf{y}}_a}}}\frac{{\partial {{\bf{y}}_a}}}{{\partial {{\bf{p}}_a}}} - \frac{{\partial \phi \left( {{z_b}} \right)}}{{\partial {z_b}}}\frac{{\partial {z_b}}}{{\partial {\Psi _a}}}\frac{{\partial {\Psi _a}}}{{\partial {{\bf{y}}_a}}}\frac{{\partial {{\bf{y}}_a}}}{{\partial {{\bf{p}}_a}}}\iota } \right.\\
	\left. { - \frac{{\partial \phi \left( {{z_c}} \right)}}{{\partial {z_c}}}\frac{{\partial {z_c}}}{{\partial {\Psi _a}}}\frac{{\partial {\Psi _a}}}{{\partial {{\bf{y}}_a}}}\frac{{\partial {{\bf{y}}_a}}}{{\partial {{\bf{p}}_a}}}J - \frac{{\partial \phi \left( {{z_d}} \right)}}{{\partial {z_d}}}\frac{{\partial {z_d}}}{{\partial {\Psi _a}}}\frac{{\partial {\Psi _a}}}{{\partial {{\bf{y}}_a}}}\frac{{\partial {{\bf{y}}_a}}}{{\partial {{\bf{p}}_a}}}\kappa } \right]\\
	= {\rm{sec}}{{\rm{h}}^2}\left( {{z_a}} \right)\left[ {{{\bf{v}}_a}{\rm{sec}}{{\rm{h}}^2}\left( {{{\bf{y}}_a}} \right)} \right]e + {\rm{sec}}{{\rm{h}}^2}\left( {{z_b}} \right)\left[ {{{\bf{v}}_b}{\rm{sec}}{{\rm{h}}^2}\left( {{{\bf{y}}_a}} \right)} \right]e\iota \\
	+ {\rm{sec}}{{\rm{h}}^2}\left( {{z_c}} \right)\left[ {{{\bf{v}}_c}{\rm{sec}}{{\rm{h}}^2}\left( {{{\bf{y}}_a}} \right)} \right]eJ + {\rm{sec}}{{\rm{h}}^2}\left( {{z_d}} \right)\left[ {{{\bf{v}}_d}{\rm{sec}}{{\rm{h}}^2}\left( {{{\bf{y}}_a}} \right)} \right]e\kappa {\rm{,}}
\end{array}\]

Others in can be obtained similarly as follows.
\[\begin{array}{c}
	\frac{{\partial e{\Phi ^ * }\left[ z \right]}}{{\partial {{\bf{p}}_b}}}\iota  = {\rm{sec}}{{\rm{h}}^2}\left( {{z_a}} \right)\left[ {{{\bf{v}}_b}{\rm{sec}}{{\rm{h}}^2}\left( {{{\bf{y}}_b}} \right)} \right]e\iota  \\+ {\rm{sec}}{{\rm{h}}^2}\left( {{z_b}} \right)\left[ {{{\bf{v}}_a}{\rm{sec}}{{\rm{h}}^2}\left( {{{\bf{y}}_b}} \right)} \right]e\left( {\begin{array}{*{20}{c}}
			{{a_{11}}}& \ldots &{{a_{1n}}}\\
			\vdots & \ddots & \vdots \\
			{{a_{m1}}}& \cdots &{{a_{mn}}}
	\end{array}} \right)\\
	- {\rm{sec}}{{\rm{h}}^2}\left( {{z_c}} \right)\left[ {{{\bf{v}}_d}{\rm{sec}}{{\rm{h}}^2}\left( {{{\bf{y}}_b}} \right)} \right]e\kappa  \\- {\rm{sec}}{{\rm{h}}^2}\left( {{z_d}} \right)\left[ {{{\bf{v}}_c}{\rm{sec}}{{\rm{h}}^2}\left( {{{\bf{y}}_b}} \right)} \right]eJ{\rm{,}}
\end{array}\]

\[\begin{array}{c}
	\frac{{\partial e{\Phi ^ * }\left[ z \right]}}{{\partial {{\bf{p}}_c}}}J = {\rm{sec}}{{\rm{h}}^2}\left( {{z_a}} \right)\left[ {{{\bf{v}}_c}{\rm{sec}}{{\rm{h}}^2}\left( {{{\bf{y}}_c}} \right)} \right]eJ \\- {\rm{sec}}{{\rm{h}}^2}\left( {{z_b}} \right)\left[ {{{\bf{v}}_d}{\rm{sec}}{{\rm{h}}^2}\left( {{{\bf{y}}_c}} \right)} \right]e\kappa \\
	{\rm{ + sec}}{{\rm{h}}^2}\left( {{z_c}} \right)\left[ {{{\bf{v}}_a}{\rm{sec}}{{\rm{h}}^2}\left( {{{\bf{y}}_c}} \right)} \right]e \\- {\rm{sec}}{{\rm{h}}^2}\left( {{z_d}} \right)\left[ {{{\bf{v}}_b}{\rm{sec}}{{\rm{h}}^2}\left( {{{\bf{y}}_c}} \right)} \right]e\iota {\rm{,}}
\end{array}\]

\[\begin{array}{c}
	\frac{{\partial e{\Phi ^ * }\left[ z \right]}}{{\partial {{\bf{p}}_d}}}\kappa  = {\rm{sec}}{{\rm{h}}^2}\left( {{z_a}} \right)\left[ {{{\bf{v}}_d}{\rm{sec}}{{\rm{h}}^2}\left( {{{\bf{y}}_d}} \right)} \right]e\kappa  \\- {\rm{sec}}{{\rm{h}}^2}\left( {{z_b}} \right)\left[ {{{\bf{v}}_c}{\rm{sec}}{{\rm{h}}^2}\left( {{{\bf{y}}_d}} \right)} \right]eJ\\
	- {\rm{sec}}{{\rm{h}}^2}\left( {{z_c}} \right)\left[ {{{\bf{v}}_b}{\rm{sec}}{{\rm{h}}^2}\left( {{{\bf{y}}_d}} \right)} \right]e\iota  \\+ {\rm{sec}}{{\rm{h}}^2}\left( {{z_d}} \right)\left[ {{{\bf{v}}_a}{\rm{sec}}{{\rm{h}}^2}\left( {{{\bf{y}}_d}} \right)} \right]e{\rm{,}}
\end{array}\]

\[\begin{array}{c}
	{\kern 1pt} {\kern 1pt}  - {\rm{4}}\frac{{\partial J}}{{\partial {{\bf{p}}^ * }}} =  - \left( {\frac{{\partial J}}{{\partial {{\bf{p}}_a}}} + \frac{{\partial J}}{{\partial {{\bf{p}}_b}}}\iota  + \frac{{\partial J}}{{\partial {{\bf{p}}_c}}}J + \frac{{\partial J}}{{\partial {{\bf{p}}_d}}}\kappa } \right)\\
	= {\rm{2}}\left[ {{\rm{sec}}{{\rm{h}}^2}\left( {{z_a}} \right){e_a}{{\bf{v}}_a} - {\rm{sec}}{{\rm{h}}^2}\left( {{z_b}} \right){e_b}{{\bf{v}}_b}} \right.\\
	\left. { - {\rm{sec}}{{\rm{h}}^2}\left( {{z_c}} \right){e_c}{{\bf{v}}_c} - {\rm{sec}}{{\rm{h}}^2}\left( {{z_d}} \right){e_d}{{\bf{v}}_d}} \right]{\rm{sec}}{{\rm{h}}^2}\left( {{{\bf{y}}_a}} \right)\\
	+ {\rm{2}}\left[ {{\rm{sec}}{{\rm{h}}^2}\left( {{z_a}} \right){e_a}{{\bf{v}}_b} + {\rm{sec}}{{\rm{h}}^2}\left( {{z_b}} \right){e_b}{{\bf{v}}_a}} \right.\\
	\left. { - {\rm{sec}}{{\rm{h}}^2}\left( {{z_c}} \right){e_c}{{\bf{v}}_d} + {\rm{sec}}{{\rm{h}}^2}\left( {{z_d}} \right){e_d}{{\bf{v}}_c}} \right]{\rm{sec}}{{\rm{h}}^2}\left( {{{\bf{y}}_b}} \right)\iota \\
	{\rm{ + 2}}\left[ {{\rm{sec}}{{\rm{h}}^2}\left( {{z_a}} \right){e_a}{{\bf{v}}_c} + {\rm{sec}}{{\rm{h}}^2}\left( {{z_b}} \right){e_b}{{\bf{v}}_d}} \right.\\
	\left. {{\rm{ + sec}}{{\rm{h}}^2}\left( {{z_c}} \right){e_c}{{\bf{v}}_a} - {\rm{sec}}{{\rm{h}}^2}\left( {{z_d}} \right){e_d}{{\bf{v}}_b}} \right]{\rm{sec}}{{\rm{h}}^2}\left( {{{\bf{y}}_c}} \right)J\\
	{\rm{ + 2}}\left[ {{\rm{sec}}{{\rm{h}}^2}\left( {{z_a}} \right){e_a}{{\bf{v}}_d} - {\rm{sec}}{{\rm{h}}^2}\left( {{z_b}} \right){e_b}{{\bf{v}}_c}} \right.\\
	\left. { + {\rm{sec}}{{\rm{h}}^2}\left( {{z_c}} \right){e_c}{{\bf{v}}_b} + {\rm{sec}}{{\rm{h}}^2}\left( {{z_d}} \right){e_d}{{\bf{v}}_a}} \right]{\rm{sec}}{{\rm{h}}^2}\left( {{{\bf{y}}_d}} \right)\kappa \\
	{\rm{ = 2}}\left( {\left[ {\frac{{\partial \Phi \left( z \right)}}{{\partial {z^ * }}} \odot e} \right]{\bf{v}}} \right) \odot \frac{{\partial \Psi \left[ {\bf{y}} \right]}}{{\partial {{\bf{y}}^{\rm{*}}}}}.
\end{array}\]

Then we have
\[{\bf{p}}\left( {i{\rm{ + 1}}} \right) = {\bf{p}}\left( i \right) + {\eta _p}\left( {{\bf{v}}\left[ {\frac{{\partial \Phi \left( z \right)}}{{\partial {z^ * }}} \odot e} \right]} \right) \odot \frac{{\partial \Psi \left( {\bf{y}} \right)}}{{\partial {{\bf{y}}^{\rm{*}}}}},\]

\subsection{Quaternion Gradient for W}

We have 
\[{\kern 1pt} {\kern 1pt} \frac{{\partial J}}{{\partial {{\bf{W}}^ * }}} = \frac{1}{4}\left( {\frac{{\partial J}}{{\partial {{\bf{W}}_a}}} + \frac{{\partial J}}{{\partial {{\bf{W}}_b}}}\iota  + \frac{{\partial J}}{{\partial {{\bf{W}}_c}}}J + \frac{{\partial J}}{{\partial {{\bf{W}}_d}}}\kappa } \right),\]

where
\[\left\{ \begin{array}{l}
	\frac{{\partial J}}{{\partial {{\bf{W}}_a}}} =  - \frac{{\partial e{\Phi ^ * }\left[ z \right]}}{{\partial {{\bf{W}}_a}}} - \frac{{\partial \Phi \left[ z \right]{e^ * }}}{{\partial {{\bf{W}}_a}}},\\
	\frac{{\partial J}}{{\partial {{\bf{W}}_b}}}\iota  =  - \frac{{\partial e{\Phi ^ * }\left[ z \right]}}{{\partial {{\bf{W}}_b}}}\iota  - \frac{{\partial \Phi \left[ z \right]{e^ * }}}{{\partial {{\bf{W}}_b}}}\iota ,\\
	\frac{{\partial J}}{{\partial {{\bf{W}}_c}}}J =  - \frac{{\partial e{\Phi ^ * }\left[ z \right]}}{{\partial {{\bf{W}}_c}}}J - \frac{{\partial \Phi \left[ z \right]{e^ * }}}{{\partial {{\bf{W}}_c}}}J,\\
	\frac{{\partial J}}{{\partial {{\bf{W}}_d}}}\kappa  =  - \frac{{\partial e{\Phi ^ * }\left[ z \right]}}{{\partial {{\bf{W}}_d}}}\kappa  - \frac{{\partial \Phi \left[ z \right]{e^ * }}}{{\partial {{\bf{W}}_d}}}\kappa .
\end{array} \right.\]
Note that 
\[\frac{{\partial e{\Phi ^ * }\left[ z \right]}}{{\partial {{\bf{W}}_\delta }}} = {\left( {\frac{{\partial \Phi \left[ z \right]{e^ * }}}{{\partial {{\bf{W}}_\delta }}}} \right)^ * },\delta  \in \left\{ {a,b,c,d} \right\}.\]

Because of the limitation of space, we only provide the details of ${{\partial e{\Phi ^ * }\left[ z \right]} \mathord{\left/
		{\vphantom {{\partial e{\Phi ^ * }\left[ z \right]} {\partial {{\bf{W}}_a}}}} \right.
		\kern-\nulldelimiterspace} {\partial {{\bf{W}}_a}}}$ : 

\[\frac{{\partial e\Phi {{\left[ z \right]}^ * }}}{{\partial {{\bf{W}}_a}}} = e\left[ {\frac{{\partial \phi \left( {{z_a}} \right)}}{{\partial {{\bf{W}}_a}}} - \frac{{\partial \phi \left( {{z_b}} \right)}}{{\partial {{\bf{W}}_a}}}\iota  - \frac{{\partial \phi \left( {{z_c}} \right)}}{{\partial {{\bf{W}}_a}}}J - \frac{{\partial \phi \left( {{z_d}} \right)}}{{\partial {{\bf{W}}_a}}}\kappa } \right],\]
where

\[\begin{array}{c}
	\frac{{\partial \phi \left( {{z_a}} \right)}}{{\partial {{\bf{W}}_a}}} = \frac{{\partial \phi \left( {{z_a}} \right)}}{{\partial {z_a}}}\frac{{\partial {z_a}}}{{\partial {\Psi _a}}}\frac{{\partial {\Psi _a}}}{{\partial {{\bf{y}}_a}}}\frac{{\partial {{\bf{y}}_a}}}{{\partial {{\bf{W}}_a}}} + \frac{{\partial \phi \left( {{z_a}} \right)}}{{\partial {z_a}}}\frac{{\partial {z_a}}}{{\partial {\Psi _b}}}\frac{{\partial {\Psi _b}}}{{\partial {{\bf{y}}_b}}}\frac{{\partial {{\bf{y}}_b}}}{{\partial {{\bf{W}}_a}}}\\
	+ \frac{{\partial \phi \left( {{z_a}} \right)}}{{\partial {z_a}}}\frac{{\partial {z_a}}}{{\partial {\Psi _c}}}\frac{{\partial {\Psi _c}}}{{\partial {{\bf{y}}_c}}}\frac{{\partial {{\bf{y}}_c}}}{{\partial {{\bf{W}}_a}}} + \frac{{\partial \phi \left( {{z_a}} \right)}}{{\partial {z_a}}}\frac{{\partial {z_a}}}{{\partial {\Psi _d}}}\frac{{\partial {\Psi _d}}}{{\partial {{\bf{y}}_d}}}\frac{{\partial {{\bf{y}}_d}}}{{\partial {{\bf{W}}_a}}}\\
	= {\rm{sec}}{{\rm{h}}^2}\left( {{z_a}} \right)\left( {{{\bf{x}}_a}{{\left[ {{{\bf{v}}_a}{\rm{sec}}{{\rm{h}}^2}\left( {{{\bf{y}}_a}} \right)} \right]}^T} + {{\bf{x}}_b}{{\left[ {{{\bf{v}}_b}{\rm{sec}}{{\rm{h}}^2}\left( {{{\bf{y}}_b}} \right)} \right]}^T}} \right.\\
	\left. { + {{\bf{x}}_c}{{\left[ {{{\bf{v}}_c}{\rm{sec}}{{\rm{h}}^2}\left( {{{\bf{y}}_c}} \right)} \right]}^T} + {{\bf{x}}_d}{{\left[ {{{\bf{v}}_d}{\rm{sec}}{{\rm{h}}^2}\left( {{{\bf{y}}_d}} \right)} \right]}^T}} \right),
\end{array}\]

\[\begin{array}{c}
	\frac{{\partial \phi \left( {{z_b}} \right)}}{{\partial {{\bf{W}}_a}}} = {\rm{sec}}{{\rm{h}}^2}\left( {{z_b}} \right)\left( {{{\bf{x}}_a}{{\left[ { - {{\bf{v}}_b}{\rm{sec}}{{\rm{h}}^2}\left( {{{\bf{y}}_a}} \right)} \right]}^T} + {{\bf{x}}_b}{{\left[ {{{\bf{v}}_a}{\rm{sec}}{{\rm{h}}^2}\left( {{{\bf{y}}_b}} \right)} \right]}^T}} \right.\\
	\left. { + {{\bf{x}}_c}{{\left[ {{{\bf{v}}_d}{\rm{sec}}{{\rm{h}}^2}\left( {{{\bf{y}}_c}} \right)} \right]}^T} + {{\bf{x}}_d}{{\left[ { - {{\bf{v}}_c}{\rm{sec}}{{\rm{h}}^2}\left( {{{\bf{y}}_d}} \right)} \right]}^T}} \right),
\end{array}\]

\[\begin{array}{c}
	\frac{{\partial \phi \left( {{z_c}} \right)}}{{\partial {{\bf{W}}_a}}} = {\rm{sec}}{{\rm{h}}^2}\left( {{z_c}} \right)\left( {{{\bf{x}}_a}{{\left[ { - {{\bf{v}}_c}{\rm{sec}}{{\rm{h}}^2}\left( {{{\bf{y}}_a}} \right)} \right]}^T} + {{\bf{x}}_b}{{\left[ { - {{\bf{v}}_d}{\rm{sec}}{{\rm{h}}^2}\left( {{{\bf{y}}_b}} \right)} \right]}^T}} \right.\\
	\left. { + {{\bf{x}}_c}{{\left[ {{{\bf{v}}_a}{\rm{sec}}{{\rm{h}}^2}\left( {{{\bf{y}}_c}} \right)} \right]}^T} + {{\bf{x}}_d}{{\left[ {{{\bf{v}}_b}{\rm{sec}}{{\rm{h}}^2}\left( {{{\bf{y}}_d}} \right)} \right]}^T}} \right),
\end{array}\]

\[\begin{array}{c}
	\frac{{\partial \phi \left( {{z_d}} \right)}}{{\partial {{\bf{W}}_a}}} = {\rm{sec}}{{\rm{h}}^2}\left( {{z_d}} \right)\left( {{{\bf{x}}_a}{{\left[ { - {{\bf{v}}_d}{\rm{sec}}{{\rm{h}}^2}\left( {{{\bf{y}}_a}} \right)} \right]}^T} + {{\bf{x}}_b}{{\left[ {{{\bf{v}}_c}{\rm{sec}}{{\rm{h}}^2}\left( {{{\bf{y}}_b}} \right)} \right]}^T}} \right.\\
	\left. { + {{\bf{x}}_c}{{\left[ { - {{\bf{v}}_b}{\rm{sec}}{{\rm{h}}^2}\left( {{{\bf{y}}_c}} \right)} \right]}^T} + {{\bf{x}}_d}{{\left[ {{{\bf{v}}_a}{\rm{sec}}{{\rm{h}}^2}\left( {{{\bf{y}}_d}} \right)} \right]}^T}} \right).
\end{array}\]

Others in can be obtained similarly. Then the gradient of W can be obtained as follows:
\[\begin{array}{l}
	- 4{\kern 1pt} {\kern 1pt} \frac{{\partial J}}{{\partial {{\bf{W}}^ * }}} = \left( {\frac{{\partial J}}{{\partial {{\bf{W}}_a}}} + \frac{{\partial J}}{{\partial {{\bf{W}}_b}}}\iota  + \frac{{\partial J}}{{\partial {{\bf{W}}_c}}}J + \frac{{\partial J}}{{\partial {{\bf{W}}_d}}}\kappa } \right)\\
	= {\rm{2}}{\bf{x}}{\left( {\left( {v\left[ {\frac{{\partial \Phi \left( z \right)}}{{\partial {z^ * }}} \odot e} \right]} \right) \odot \left( {\frac{{\partial \Psi \left[ {\bf{y}} \right]}}{{\partial {{\bf{y}}^{\rm{*}}}}}} \right)} \right)^H}.
\end{array}\]

Finally, the quaternion LMS algorithm for the MLP (19) with the split quaternion activation function is summarized as

\[{\kern 1pt} {\kern 1pt} q\left( {i{\rm{ + 1}}} \right) = q\left( i \right) + {\eta _q}\frac{{\partial \Phi \left( z \right)}}{{\partial {z^ * }}} \odot e\left( i \right),\]

\[{\kern 1pt} {\bf{V}}\left( {i{\rm{ + 1}}} \right) = {\bf{V}}\left( i \right) + {\eta _v}\Psi \left( {\bf{y}} \right)\left[ {{\kern 1pt} {\kern 1pt} \frac{{\partial \Phi \left( z \right)}}{{\partial {z^ * }}} \odot {e^ * }\left( i \right)} \right],\]

\[{\bf{p}}\left( {i{\rm{ + 1}}} \right) = {\bf{p}}\left( i \right) + {\eta _p}\left( {{\bf{v}}\left[ {\frac{{\partial \Phi \left( z \right)}}{{\partial {z^ * }}} \odot e} \right]} \right) \odot \frac{{\partial \Psi \left( {\bf{y}} \right)}}{{\partial {{\bf{y}}^{\rm{*}}}}},\]

\[{\bf{W}}\left( {i{\rm{ + 1}}} \right) = {\bf{W}}\left( i \right) + {\eta _w}{\bf{x}}{\left[ {\left( {v\left[ {\frac{{\partial \Phi \left( z \right)}}{{\partial {z^ * }}} \odot e} \right]} \right) \odot \left( {\frac{{\partial \Psi \left( {\bf{y}} \right)}}{{\partial {{\bf{y}}^{\rm{*}}}}}} \right)} \right]^H}.\]

The obtained algorithm (38a–d) has the similar expression to that in the real and complex domains.	

\section{Quaternion MLP Based on The MCC}
The cost function of the MCC is defined as 

\[{J_{MCC}}\left( {\bf{w}} \right) = \exp \left( { - \frac{{e(i)e_{}^ * (i)}}{{2{\sigma ^2}}}} \right),{\kern 1pt} \]

to estimate the parameter ${\bf{w}} \in {H^n}$ . Then, the corresponding quaternion stochastic gradient ascent algorithm for the MLP based on the MCC can be derived from (38a–d) as
\[{\kern 1pt} {\kern 1pt} q\left( {i{\rm{ + 1}}} \right) = q\left( i \right) + {\eta _q}{J_{MCC}}\left( {\bf{w}} \right)\frac{{\partial \Phi \left( z \right)}}{{\partial {z^ * }}} \odot e\left( i \right),\]

\[{\kern 1pt} {\bf{V}}\left( {i{\rm{ + 1}}} \right) = {\bf{V}}\left( i \right) + {\eta _v}{J_{MCC}}\left( {\bf{w}} \right)\Psi \left( {\bf{y}} \right)\left[ {{\kern 1pt} {\kern 1pt} \frac{{\partial \Phi \left( z \right)}}{{\partial {z^ * }}} \odot {e^ * }\left( i \right)} \right],\]

\[{\bf{p}}\left( {i{\rm{ + 1}}} \right) = {\bf{p}}\left( i \right) + {\eta _p}{J_{MCC}}\left( {\bf{w}} \right)\left( {{\bf{v}}\left[ {\frac{{\partial \Phi \left( z \right)}}{{\partial {z^ * }}} \odot e} \right]} \right) \odot \frac{{\partial \Psi \left( {\bf{y}} \right)}}{{\partial {{\bf{y}}^{\rm{*}}}}},\]

\[{\bf{W}}\left( {i{\rm{ + 1}}} \right) = {\bf{W}}\left( i \right) + {\eta _w}{J_{MCC}}\left( {\bf{w}} \right){\bf{x}}{\left[ {\left( {v\left[ {\frac{{\partial \Phi \left( z \right)}}{{\partial {z^ * }}} \odot e} \right]} \right) \odot \left( {\frac{{\partial \Psi \left( {\bf{y}} \right)}}{{\partial {{\bf{y}}^{\rm{*}}}}}} \right)} \right]^H}.\]

\section{Simulation}

In the simulations, we use two proposed quaternion MLP algorithms to predict the chaotic time-series, which were generated using the following Mackey–Glass time-delay differential equation.

\[\dot x\left( t \right) = \frac{{0.2x\left( {t - \tau } \right)}}{{1 + {x^{10}}\left( {t - \tau } \right)}} - 0.1x\left( t \right),\]

where $\tau  = 17$, $x\left( 0 \right){\rm{ = 0}}{\rm{.12}}$, and  for all $t < {\rm{0}}$ . In this prediction task, we used the previous five historical data to predict the next data in the future. Thus, the sizes of the inputs and outputs were  and , the parameter in the hidden layer had dimensions of  , and the size of  was  . 
\subsection{Prediction without Noises}

First, we simulated the one–step prediction of Mackey–Glass data without any noise. We set the step size  . Fig. 1 shows that the MSE algorithm and MCC algorithm obtained similar results, so we only present the good prediction results of the MSE algorithm in Fig. 2. The prediction results include one real part and three imaginary components.

\begin{figure}[!t]
	\includegraphics[width=\columnwidth]{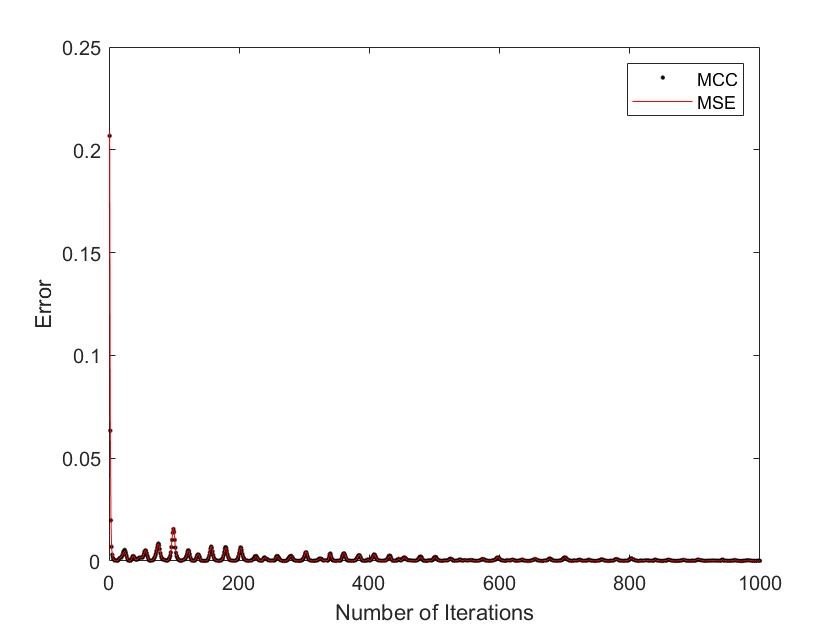}
	\caption{Error comparison between the MCC and MSE without noise}
	\label{fig2}
\end{figure}
\begin{figure}[!t]
	\includegraphics[width=\columnwidth]{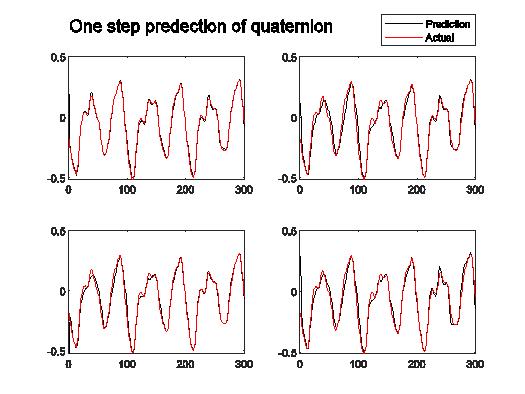}
	\caption{One–step prediction results for the quaternion MSE without noise.}
	\label{fig3}
\end{figure}

\subsection{Prediction under Noise}
Additionally, we assumed that the Mackey–Glass data are noisy. To simulate the representative noise scenario, two types of noise signals were generated. 
The first type of noise was Gaussian. We set the step size  ; the simulations of the MCC and MSE algorithms are shown in Fig. 3. From the comparison, we found that the MCC algorithm converged slower than the MSE algorithm while the steady errors of MCC performs better than MSE.The difference between the steady-state errors of the two algorithms is more obvious in Figure 3 (b) in dB.

\begin{figure}[!t]
	\includegraphics[width=\columnwidth]{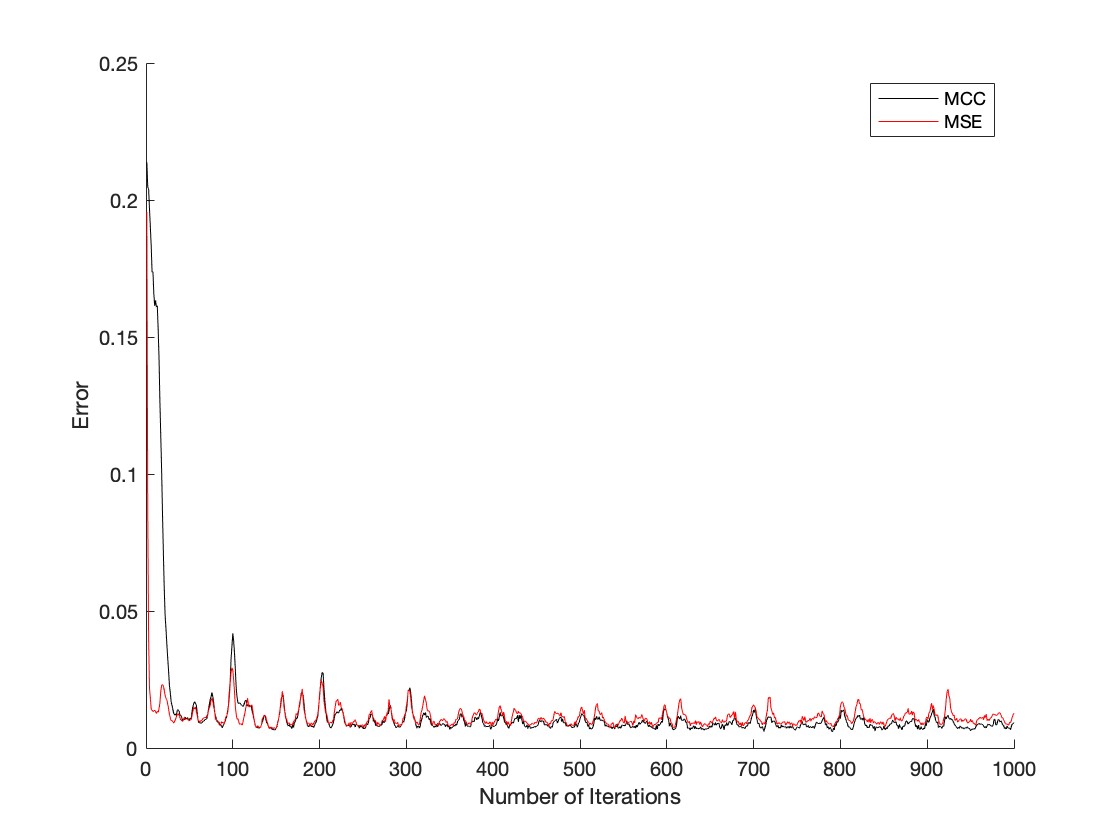}
	\caption{Error comparison under Gaussian noise}
	\label{fig4}
\end{figure}
\begin{figure}[!t]
	\includegraphics[width=\columnwidth]{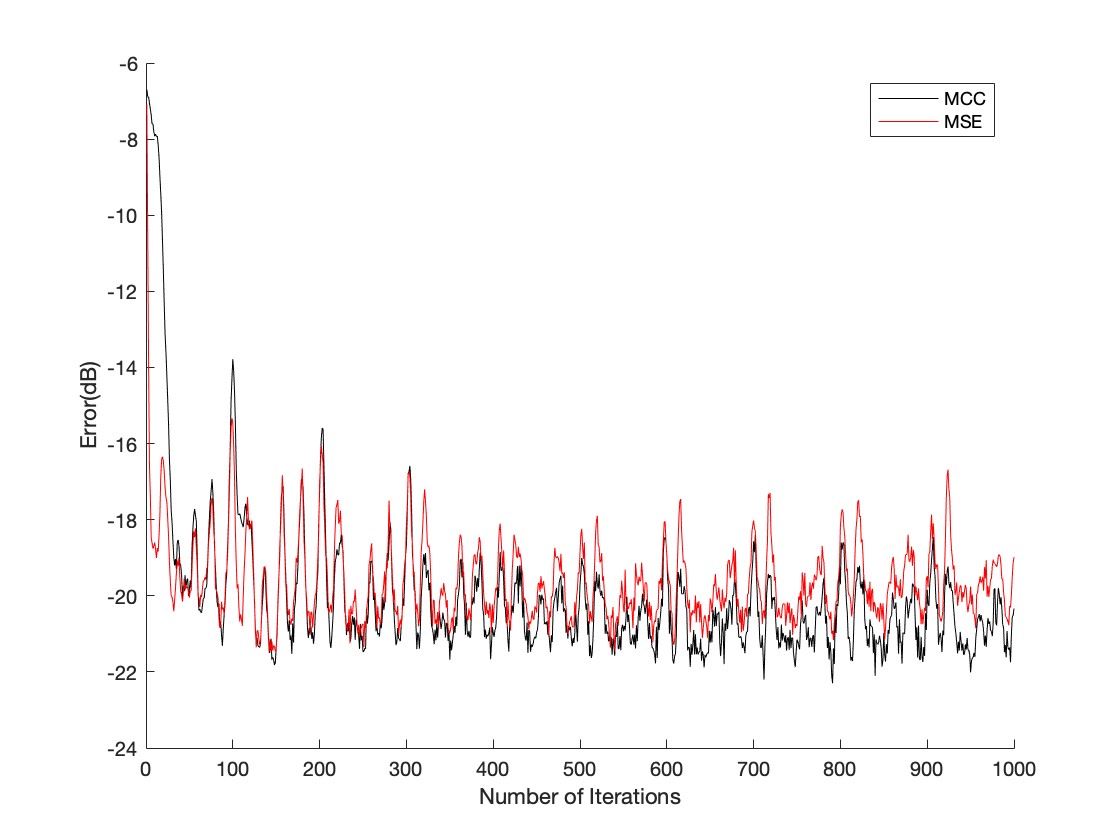}
	\caption{Error comparison under Gaussian noise}
	\label{fig5}
\end{figure}
\begin{figure}[!t]
	\includegraphics[width=\columnwidth]{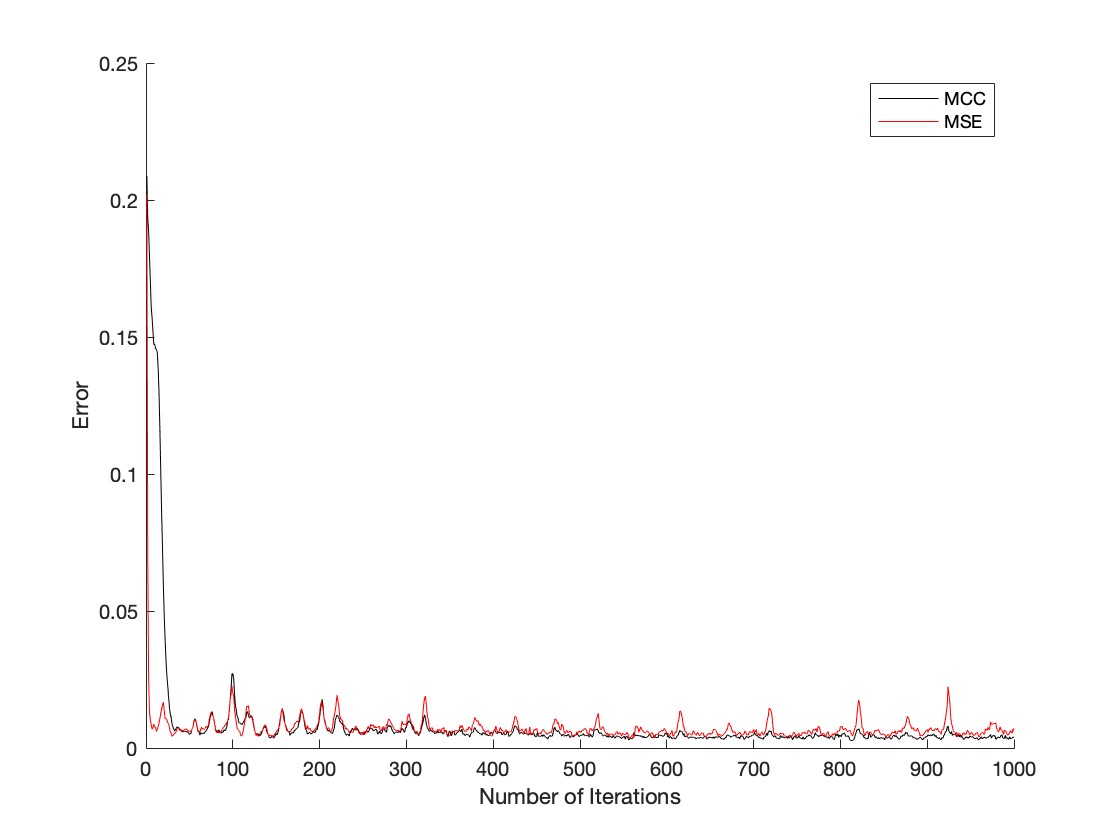}
	\caption{ Error comparison under impulsive noise}
	\label{fig6}
\end{figure}
\begin{figure}[!t]
	\includegraphics[width=\columnwidth]{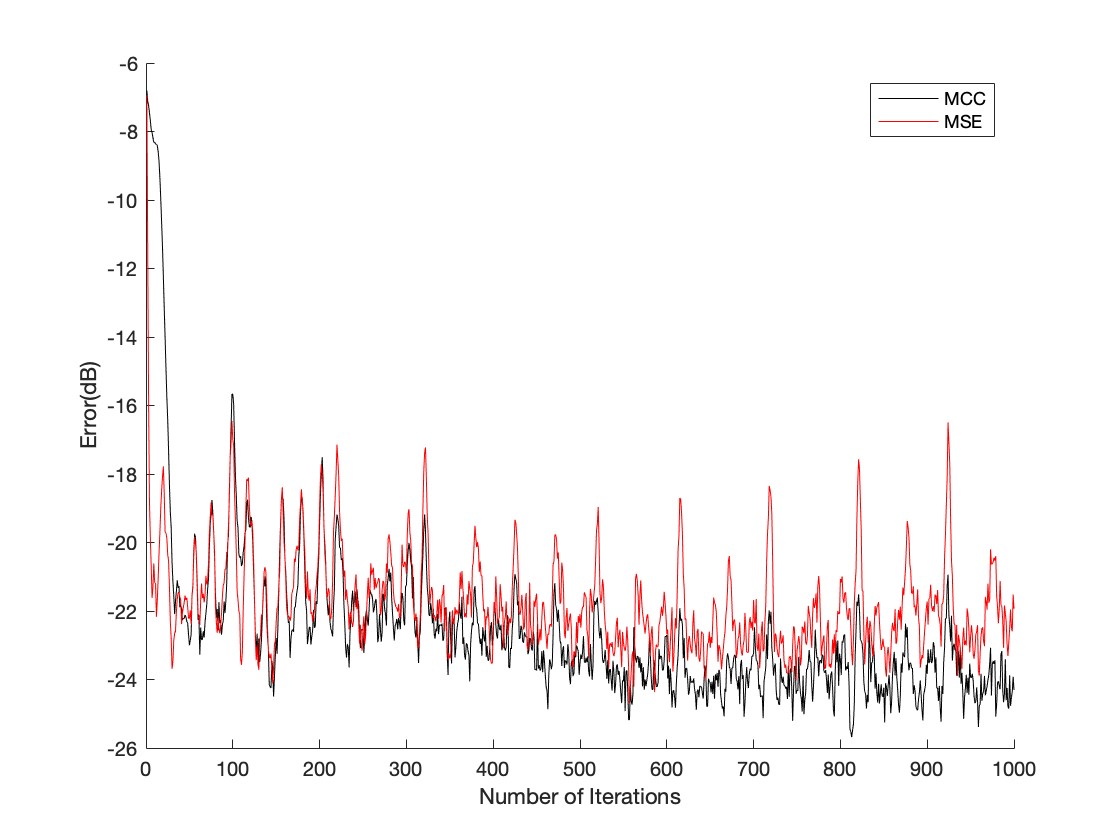}
	\caption{ Error comparison under impulsive noise}
	\label{fig7}
\end{figure}

\section{Conclusion}
In this paper, we have proposed two algorithms for quaternion MLP networks based on the cost functions of the MSE and MCC, respectively. In the algorithms, we utilized the GHR quaternion gradient. The two algorithms performed similarly under noiseless and Gaussian noise circumstances. The MCC algorithm performed better than the MSE algorithm under impulsive noise.


\begin{thebibliography}{}
\providecommand{\url}[1]{#1}
\csname url@samestyle\endcsname
\providecommand{\newblock}{\relax}
\providecommand{\bibinfo}[2]{#2}
\providecommand{\BIBentrySTDinterwordspacing}{\spaceskip=0pt\relax}
\providecommand{\BIBentryALTinterwordstretchfactor}{4}
\providecommand{\BIBentryALTinterwordspacing}{\spaceskip=\fontdimen2\font plus
\BIBentryALTinterwordstretchfactor\fontdimen3\font minus
  \fontdimen4\font\relax}
\providecommand{\BIBforeignlanguage}[2]{{%
\expandafter\ifx\csname l@#1\endcsname\relax
\typeout{** WARNING: IEEEtran.bst: No hyphenation pattern has been}%
\typeout{** loaded for the language `#1'. Using the pattern for}%
\typeout{** the default language instead.}%
\else
\language=\csname l@#1\endcsname
\fi
#2}}
\providecommand{\BIBdecl}{\relax}
\BIBdecl

\end{thebibliography}


\section*{References}
\begin{thebibliography}{99}
\bibitem{bib1}1. Y. Liu, D. Zhang, J. Lou, J. Lu and J. Cao, "Stability Analysis of Quaternion-Valued Neural Networks: Decomposition and Direct Approaches," in IEEE Transactions on Neural Networks and Learning Systems, vol. 29, no. 9, pp. 4201-4211, Sept. 2018.
\bibitem{bib2} X. Chen, Q. Song, Z. Li, Z. Zhao and Y. Liu, "Stability Analysis of Continuous-Time and Discrete-Time Quaternion-Valued Neural Networks With Linear Threshold Neurons," in IEEE Transactions on Neural Networks and Learning Systems, vol. 29, no. 7, pp. 2769-2781, July 2018.
\bibitem{bib3} X. Yang, C. Li, Q. Song, H. Li and J. Huang, "Effects of State-Dependent Impulses on Robust Exponential Stability of Quaternion-Valued Neural Networks Under Parametric Uncertainty," in IEEE Transactions on Neural Networks and Learning Systems, vol. 30, no. 7, pp. 2197-2211, July 2019.
\bibitem{bib4} X. Fan, G. Wang, J. Han, Y. Wang, A Background-Impulse Kalman Filter with Non-Gaussian Measurement Noises, IEEE Transactions on Systems, Man and Cybernetics: Systems, vol. 53, no. 4, pp. 2434-2443, April. 2023.
\bibitem{bib5} M. E. Valle and F. Z. de Castro, "On the Dynamics of Hopfield Neural Networks on Unit Quaternions," in IEEE Transactions on Neural Networks and Learning Systems, vol. 29, no. 6, pp. 2464-2471, June 2018.
\bibitem{bib6} X. Chen, Q. Song and Z. Li, "Design and Analysis of Quaternion-Valued Neural Networks for Associative Memories," in IEEE Transactions on Systems, Man, and Cybernetics: Systems, vol. 48, no. 12, pp. 2305-2314, Dec. 2018.
\bibitem{bib7} Zhenyu Feng , Gang Wang , Bei Peng* , Jiacheng He , Kun Zhang , Distributed Minimum Error Entropy Kalman Filter, Information Fusion (2023), Volume 91, March 2023, Pages 556-565.
\bibitem{bib8} V. Risojević and Z. Babić, "Unsupervised Quaternion Feature Learning for Remote Sensing Image Classification," in IEEE Journal of Selected Topics in Applied Earth Observations and Remote Sensing, vol. 9, no. 4, pp. 1521-1531, April 2016.
\bibitem{bib9}Q. Yin, J. Wang, X. Luo, J. Zhai, S. K. Jha and Y. Shi, "Quaternion Convolutional Neural Network for Color Image Classification and Forensics," in IEEE Access, vol. 7, pp. 20293-20301, 2019.
\bibitem{bib10} JiachengHe, Gang Wang*, Kui Cao, He Diao, Guotai Wang, Bei Peng, Generalized minimum error entropy robust learning, Pattern Recognition, 2023, vol. 135, pp. 109188.
\bibitem{bib11} H. Kim and A. Hirose, "Unsupervised Fine Land Classification Using Quaternion Autoencoder-Based Polarization Feature Extraction and Self-Organizing Mapping," in IEEE Transactions on Geoscience and Remote Sensing, vol. 56, no. 3, pp. 1839-1851, March 2018.
\bibitem{bib12} D. Zheng, H. Zhang, J. A. Zhang, G. Wang, Consensus of multi-agent systems with faults and mismatches under switched topologies using a delta operator method, NEUROCOMPUTING, Vol. 315, Pages 198-209, 2018
\bibitem{bib13} Z. H. He, Q. W. Wang and Y. Zhang, “A system of quaternary coupled Sylvester-type real quaternion matrix equations,” Automatica, vol. 87, pp. 25-31, 2018.
\bibitem{bib14} G. Wang, R. Xue, J. Zhao, “Switching Criterion for Sub- and Super-Gaussian Additive Noise in Adaptive Filtering,” Signal Processing, Volume 150, September 2018, Pages 166–170.
\bibitem{bib15} B. C. Ujang, C. C. Took, and D. P. Mandic, “A split quaternion nonlinear adaptive filtering,” Neural Networks, vol. 23, no. 3, pp, 426-434, 2010.
\bibitem{bib16}B. C. Ujang, C. C. Took and D. P. Mandic, "Quaternion-Valued Nonlinear Adaptive Filtering," in IEEE Transactions on Neural Networks, vol. 22, no. 8, pp. 1193-1206, Aug. 2011.
\bibitem{bib17}G. Wang, R. Xue, Z. Chao, G. Junjie, “Complex-Valued Adaptive Networks Based on Entropy Estimation,” Signal Processing, Volume 149, August 2018, Pages 124-134.
\bibitem{bib18} Y. He, X. Zhang and X. Peng, "A Model-Free Hull Deformation Measurement Method Based on Attitude Quaternion Matching," in IEEE Access, vol. 6, pp. 8864-8869, 2018.
\bibitem{bib19}J. Zhao, H. Zhang, G. Wang, Projected Kernel Recursive Maximum Correntropy, IEEE Transactions on Circuits and Systems II: Express Briefs. vol. 65, no. 7, pp. 963-967, July 2018.
\bibitem{bib20}T. Parcollet, M. Morchid and G. Linarès, "Quaternion Convolutional Neural Networks for Heterogeneous Image Processing," ICASSP 2019 - 2019 IEEE International Conference on Acoustics, Speech and Signal Processing (ICASSP), Brighton, United Kingdom, 2019, pp. 8514-8518.
\bibitem{bib21} Ye Yalan, Sheu Phillip C -Y, Zeng Jiazhi, Wang Gang, Lu Ke. An Efficient Semi-blind Source Extraction Algorithm and Its Applications to Biomedical Signal Extraction. Science in China, Series F: Information Sciences. 2009, 52(10): 1863-1874
\bibitem{bib22} Parcollet Titouan, Zhang Ying, Morchid Mohamed,Trabelsi Chiheb, Linarès Georges, De Mori Renato, and Bengio Yoshua, “Quaternion convolutional neural networks for end-to-end automatic speech recognition,” arXiv preprint arXiv:1806.07789, 2018.
\bibitem{bib23}H. Zhang and H. Lv, "Augmented Quaternion Extreme Learning Machine," in IEEE Access, vol. 7, pp. 90842-90850, 2019.
\bibitem{bib24}Gang Wang, Rui Xue, Quaternion Filtering Based on Quaternion Involutions and Its Application in Signal Processing, IEEE Access, Volume: 7, Issue:1, pp. 149068-149079, 2019
\bibitem{bib25} G. Huang, G.-B. Huang, S. Song, K. You, “Trends in extreme learning machines: a review,” Neural Netw. vol. 61, pp. 32–48, 2015.
\bibitem{bib26}D. Xu, Y. Xia and D. P. Mandic, "Optimization in Quaternion Dynamic Systems: Gradient, Hessian, and Learning Algorithms," in IEEE Transactions on Neural Networks and Learning Systems, vol. 27, no. 2, pp. 249-261, Feb. 2016.
\bibitem{bib27} Gang Wang, Shuizhi Sam Ge, Rui Xue, Ji Zhao, Chao Li, “Complex-valued Kalman filters based on Gaussian entropy,” Signal Processing, Volume 160, July 2019, pp. 178-189.
\bibitem{bib28} N. Benvenuto and F. Piazza, "On the complex backpropagation algorithm," in IEEE Transactions on Signal Processing, vol. 40, no. 4, pp. 967-969, April 1992.
\bibitem{bib29}G. M. Georgiou and C. Koutsougeras, "Complex domain backpropagation," in IEEE Transactions on Circuits and Systems II: Analog and Digital Signal Processing, vol. 39, no. 5, pp. 330-334, May 1992.
\bibitem{bib30} D. Xu, and D. P. Mandic, “The Theory of Quaternion Matrix Derivatives,” IEEE Trans. Signal Processing, vol. 63, no. 6, pp. 1543-1556, 2015.
\bibitem{bib31} M. Xiang, S. Kanna, and D. P. Mandic. "Performance Analysis of Quaternion-valued Adaptive Filters in Non-stationary Environments." IEEE Trans. Signal Processing, vol. 66, no. 6, pp. 1566-1579, 2018.
\bibitem{bib32}Wang Gang, Rao Nini. A Fragile Watermarking Scheme for Medical Image. 27th Annual International Conference of the IEEE Engineering in Medi
\bibitem{bib33}G. Wang, R. Xue, Comments on “The Quaternion LMS Algorithm for Adaptive Filtering of Hypercomplex Processes”, IEEE Trans. Signal Processing, vol. 67, no. 7, pp. 1957-1958, 2019.
\bibitem{bib34}  E. Hitzer, “Algebraic foundations of split hypercomplex nonlinear adaptive filtering,” Mathematical Methods in the Applied Sciences, vol. 36, no. 9, pp. 1042-1055, 2013.
\bibitem{bib35}S. Battilotti, F. Cacace, M. d’Angelo and A. Germani, "The Polynomial Approach to the LQ Non-Gaussian Regulator Problem Through Output Injection," in IEEE Transactions on Automatic Control, vol. 64, no. 2, pp. 538-552, Feb. 2019.
\bibitem{bib36} G. Wang, R. Xue, J. Zhao, “Switching Criterion for Sub- and Super-Gaussian Additive Noise in Adaptive Filtering,” Signal Processing, vol. 150, pp. 166–170, 2018.
\bibitem{bib37} G. Wang, R. Xue*, J. Wang, “A Distributed Maximum Correntropy Kalman Filter,” Signal Processing, 2019, Signal Processing, vol. 160, pp. 247-251, 2019.
\bibitem{bib38}  C. Safarian, T. Ogunfunmi, “The quaternion minimum error entropy algorithm with fiducial point for nonlinear adaptive systems,” Signal Processing, vol. 163, pp. 188-200, 2019. 
\end{thebibliography}
\end{document}